\documentclass[preprint2, floatfix,final]{aastex}

\usepackage{graphicx}
\usepackage{subfigure}

\usepackage{natbib}

\begin{document}


\title{Suppression of energetic electron transport in flares by double layers}


 
\author{T.C. Li, J.F. Drake and M. Swisdak} 
 
\affil{IREAP, University of Maryland, College Park, Maryland 20742, USA }


\date{ \today}

\begin{abstract}
During flares and coronal mass ejections, energetic electrons from coronal sources typically have very long lifetimes compared to the transit times across the systems, suggesting confinement in the source region. Particle-in-cell simulations are carried out to explore the mechanisms of energetic electron transport from the corona to the chromosphere and possible confinement. We set up an initial system of pre-accelerated hot electrons in contact with ambient cold electrons along the local magnetic field, and let it evolve over time. Suppression of transport by a nonlinear, highly localized electrostatic electric field (in the form of a double layer) is observed after a short phase of free-streaming by hot electrons. The double layer (DL) emerges at the contact of the two electron populations. It is driven by an ion-electron streaming instability due to the drift of the back-streaming return current electrons interacting with the ions. The DL grows over time and supports a significant drop in temperature and hence reduces heat flux between the two regions that is sustained for the duration of the simulation. This study shows transport suppression begins when the energetic electrons start to propagate away from a coronal acceleration site. It also implies confinement of energetic electrons with kinetic energies less than the electrostatic energy of the DL for the DL lifetime, which is much longer than the electron transit time through the source region. 

\end{abstract}



\section{\label{intro}INTRODUCTION}

In flares and coronal mass ejections, coronal electrons can be accelerated above 100 keV \citep{Lin03, Krucker10, Krucker07}, i.e., more than three orders of magnitudes higher than the ambient coronal temperature of 100 eV. The energy is ultimately believed to come from magnetic fields through magnetic reconnection, which occurs throughout the corona. X-rays and microwaves produced by these energetic electrons are observed in both the chromosphere and the corona. How energetic electrons are transported from a coronal acceleration site to the chromosphere is a key issue in understanding the dynamics of flares. Recent time-of-flight measurements of hard X-ray emission during solar flares \citep{Aschwanden96} report relative time delays of arrival at the chromosphere by electrons of different energies propagating from the coronal acceleration site. The energy dependence of these time delays provides evidence that electrons can propagate along magnetic field lines by free-streaming. However, hard X-ray observations from above-the-looptop (ALT) sources \citep{Masuda94, Krucker10, Krucker07} suggest confinement of the energetic electrons above the bright soft X-ray loops for periods several orders of magnitude longer than the (free-streaming) transit time across the source. Increased detection sensitivity indicates ALT hard X-ray emission is likely a common feature of all flares \citep{Petrosian02} although there have been fewer detections of ALT emission than chromospheric emission because the higher density, and hence emission intensity, in the chromosphere overshadows ALT emission. In this case, there must be mechanisms that inhibit the transport of electron heat flux. It is puzzling why some electrons freely propagate to the chromosphere while some are trapped in the corona after being accelerated. The physics of how energetic electrons propagate from the corona to the chromosphere, and the contrary, how confinement of electrons happens, remains poorly understood.

The transport of electron heat flux has previously been modeled as conduction mediated by classical collisions \citep{Spitzer62}, convection at a subthermal characteristic speed observed in simulations of laser fusion studies \citep{Mannheimer75}, and turbulent transport limited by anomalous resistivity \citep{Manheimer77, Tsytovich71}, also called anomalous conduction. Recent analytical and numerical studies of the transport of super-hot electron fluxes with energies greater than 10 keV show that the classical conduction model is inapplicable because the mean-free-path of electrons exceeds the characteristic length scales of the system \citep{Oreshina11}.

In the turbulent transport scenario, anomalous resistivity arises from electron scattering by turbulent wave fields excited by instabilities involving the interaction of the energetic electrons, ambient electrons and/or ions. One commonly considered mode is the ion-acoustic instability \citep{Manheimer77, Tsytovich71, Smith79}. When accelerated electrons propagate outwards, ambient electrons are drawn in as a return current. Their relative drift with ions excites ion-acoustic waves, which scatter the energetic electrons and anomalously enhance resistivity \citep{Manheimer77}. The heat flux carried by the energetic electrons is limited by the saturation of the instability due to ion trapping.

The formation of an anomalous conduction front due to ion acoustic turbulence was considered as a means to confine hot electrons for the production of hard X-rays \citep{Smith79}. The model, based on a 1D one-fluid code with a grid spacing of 100 km ($\sim$10$^4$ ion inertial lengths), could not capture processes occurring at electron scales. However, it was argued that the front propagates at approximately the ion acoustic speed $c_s$ and, with some heuristic analysis, predicted that the front would become thinner than the grid spacing, although this was artificially disallowed to avoid numerical instability. More recently, the existence of a thermal front, literally defined as a region that links plasmas in thermal nonequilibrium and sustains the temperature difference for longer than the electron free-streaming time, was studied in 1D electrostatic Vlasov simulations \citep{Arber09}. The main goal was to refute the results of an earlier particle-in-cell (PIC) simulation \citep{McKean90} in which a conduction front was not seen. The authors observed the formation of a temperature difference propagating at a speed comparable to $c_s$ and identified the behavior as a conduction front. The physics of the responsible mechanism was not identified or investigated. We suspect that front to be an ion acoustic shock and the temperature difference to be a result of shock heating due to its extremely sharp transition. We observe similar behavior in our simulation, to be presented in section \ref{result}.

To gain a better understanding of the problem of energetic electron transport from the corona to the chromosphere, an important question that must be answered is whether thermal conduction is suppressed when the accelerated electrons start to propagate into the immediate surrounding plasma. Our goal in this paper is to answer this basic question.

To study the physics at the electron scale and the nonlinear aspects of the problem, we perform a 2D PIC simulation. We observe that the transport of energetic electrons is suppressed by a double layer that occurs at the shortest scale in a plasma, the Debye length $\lambda_{De}$.

A double layer (DL) is a nonlinear electrostatic electric field localized within two adjacent layers of equal and opposite net charge \citep{Block78,Raadu88,Singh87}. It is highly localized to $\sim$10 $\lambda_{De}$ because quasi-neutrality in a plasma can only be violated on $\lambda_{De}$ scales. It can be understood as a capacitor in a plasma. An ideal DL is a unipolar electric field, which corresponds to a monotonic drop in the potential across the structure. In most situations, however, the potential can contain dips and bumps at the high or low potential sides, but with a net potential drop across the entire layer. The potential drop $\phi$ can accelerate, decelerate and reflect particles entering from the two sides of the DL. For example, electrons incoming from the high $\phi$ side will be decelerated (or reflected) if their kinetic energy is greater (or less) than $\phi$. DLs can be classified into strong and weak depending on whether $\phi$ is much greater or comparable to the mean energy of the reflected particles on either side of the DL \citep{Raadu88}. Much research interest on DLs has focused on their ability to directly accelerate particles in auroral (ionospheric and magnetospheric) and astrophysical plasmas \citep{Singh87}. For particle acceleration, the potential across the DL is usually maintained by some external energy source. Mechanisms that have been used in simulations to produce DLs include an imposed potential drop or current \citep{Singh87}. In the latter case, DLs occur as nonlinear waves arising from current-driven instabilities. The DL observed in our simulation is generated by imposing a large field-aligned temperature jump in the initial state which results in strong currents that drive the DL.

Parameters suitable for flare settings are used in our simulations and described in section \ref{sim}. In section \ref{result}, we present the results. A brief discussion of our simulations and application to confining above-the-looptop hard X-ray sources are given in section \ref{dis_app}. We summarize in section \ref{con}.

\section{\label{sim}SIMULATION}

\begin{figure}
  \includegraphics[scale=0.3,trim=4cm 0cm 0cm 14cm, clip=true, totalheight=0.1\textheight]{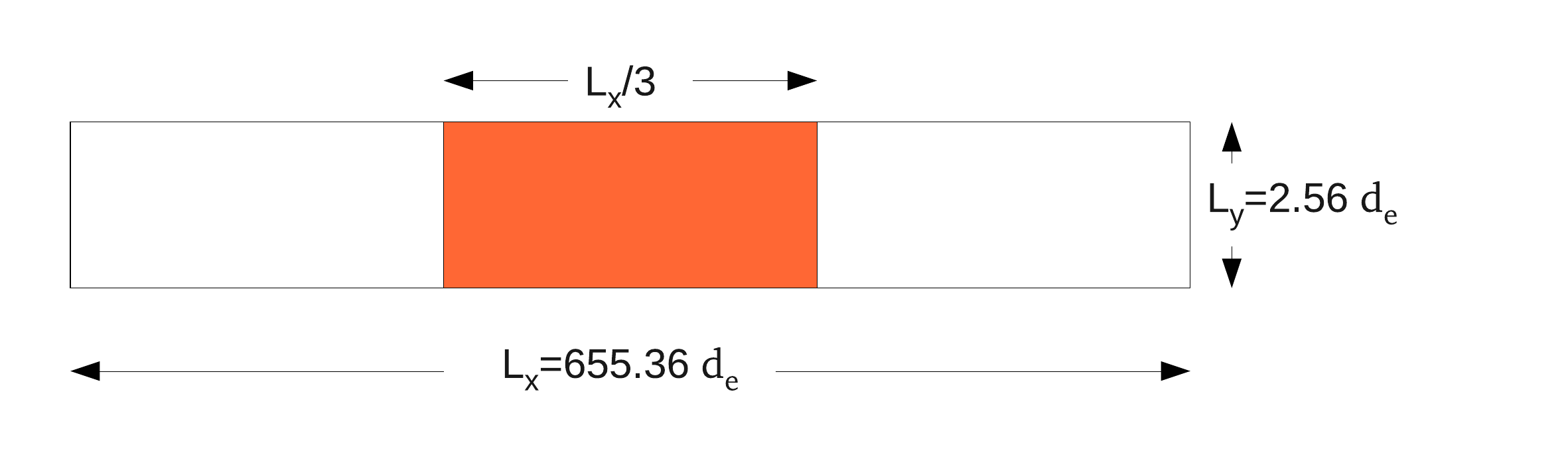}
\caption{\label{box}  A cartoon of the initial simulation setup.   }
\end{figure}

We are exploring the problem of energetic electron transport from the corona to the chromosphere by a two-dimensional electromagnetic particle-in-cell simulation using the p3d code \citep{Zeiler02}. The energetic electrons are pre-heated (i.e., present at the start of the simulation). We consider it to be more natural that the hot population not have a preferred direction of propagation, so the simulated energetic electrons are not beamed in the initial state. Fig. \ref{box} is a cartoon of the initial simulation setup. Our system represents a symmetric local segment of a flare loop with very hot electrons centered at the loop-top in contact with ambient cold electrons. The computational size is $L_x \times L_y$= 655.36 $\times$ 2.56 $d_e^2$ with a cell size of 0.02$\times$0.02 $d_e^2$. There are 400 particles per cell. $x$ is the direction parallel to the initial background magnetic field $B_0$. $d_e$=$c/\omega_{pe}$ is the electron inertial length, where $\omega_{pe}$=$(4\pi n_0e^2/m_e)^{1/2}$ is the electron plasma frequency and $n_0$ the initial density, which is uniform. The boundaries are periodic in both dimensions. The simulation domain is of course far smaller than any realistic flare loop.


The electrons are initially modeled by bi-Maxwellian distributions and the ions by a Maxwellian distribution. At the simulation's start, one-third of the electrons, centered at the middle of the domain, have temperatures parallel to the background magnetic field $T_{e,\parallel}/m_ic_A^2$=0.5 (orange in Fig. \ref{box}) while the remaining two-thirds have $T_{e,\parallel}/m_ic_A^2$=0.1, corresponding to ratios of parallel plasma pressure to magnetic pressure of $\beta_{e,\parallel}$=1 and 0.2, respectively. $c_A$=$B_0/(4\pi m_in_0)^{1/2}$ is the Alfv$\acute{\mbox{e}}$n speed based on $B_0$. The temperature perpendicular to $B_0$ is $T_{e,\perp}$=0.1=$T_i$ everywhere, so $\beta_{e,\perp}$=0.2=$\beta_i$. Since it is easier to accelerate electrons along the field lines, the hot electrons are taken to be anisotropic. Ambient electrons and ions are taken to be isotropic. Note that the unity $\beta$ hot electron pressure is consistent with recent over-the-limb flare observations \citep{Krucker10}.

Other parameters used are: mass ratio $m_e/m_i$=1/100 and speed of light $c/c_A$=100. Electric fields are normalized to $E_0$=$c_AB_0/c$. Thermal speeds are defined as $v_{tx}$=$(2T_x/m_x)^{1/2}$. The Debye length based on the initial hot electron temperature is $\lambda_{De}$=$v_{te0,h}/\omega_{pe}$=0.1 $d_e$. We will use $d_e$ as the unit of length, but it can be conveniently converted to $\lambda_{De}$. We will use $T_e$ to represent $T_{e,\parallel}$ in the following for simplicity, since it is the parallel heat transport that is of most interest. The total duration of the simulation is less than the transit time of the domain by electrons at 1.5 $v_{te0,h}$, so the majority of hot electrons will not encounter a boundary during a run.

\section{\label{result}RESULTS}

\subsection{A jump in electron temperature}

\begin{figure}
\includegraphics[scale=0.55]{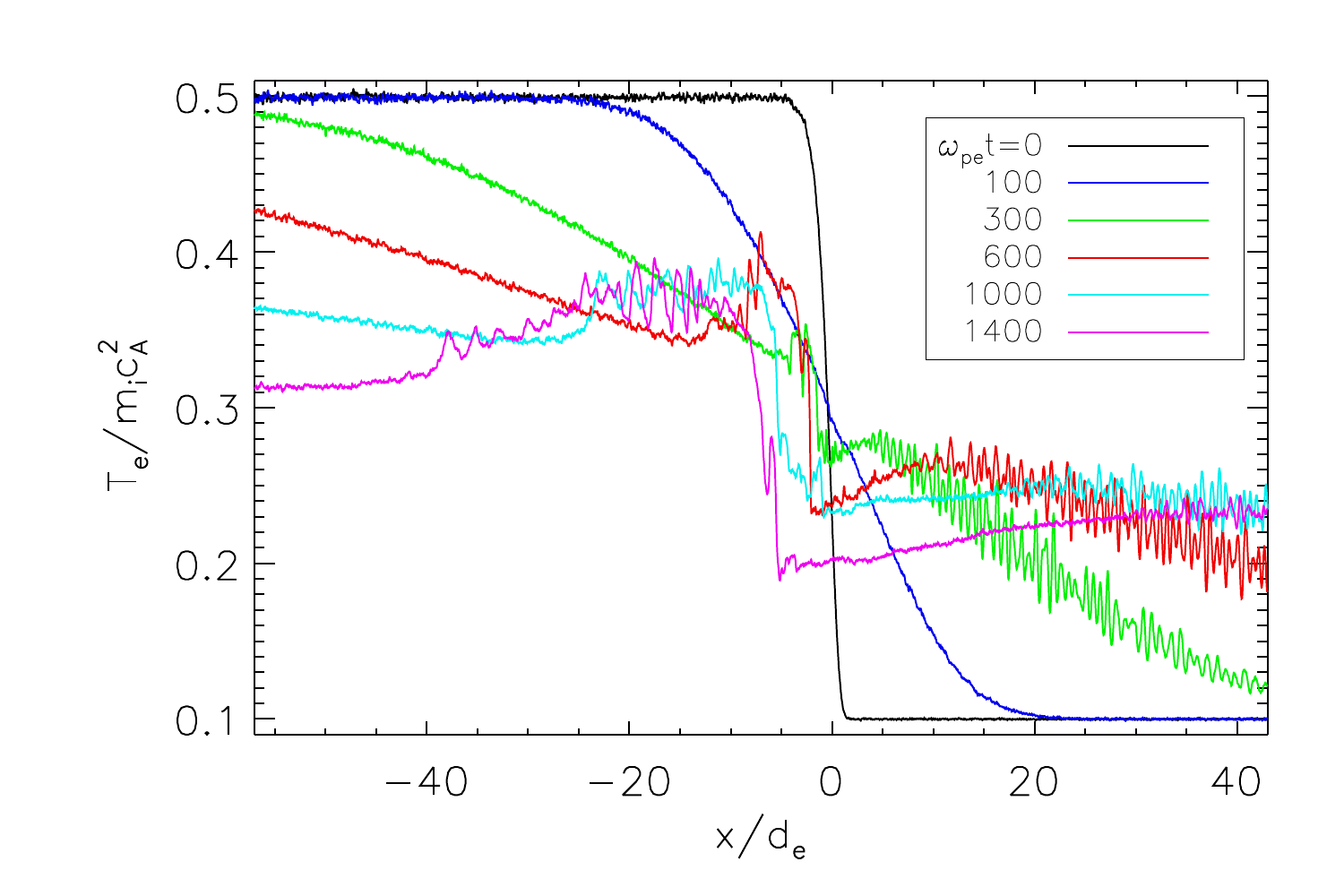}
\caption{\label{Te}  Evolution of electron temperature $T_e$ parallel to the local magnetic field near the transition between hot and cold electrons, at x= 0$d_e$. We begin with two electrons populations in contact (black) and let them evolve. The initial gradient becomes a smooth diffusive profile at $\omega_{pe}$t=100 (blue). The smooth transition is then interrupted by a jump which signifies transport suppression from the high to low $T_e$ sides.    }
\end{figure}

The primary drive of the evolution of the system is the contact of hot electrons with cold electrons, so the dynamics starts in the region of contact. Since this is a symmetric system, it suffices to study either side of the contact. We choose the right side, where hot electrons propagate outwards with positive velocities. The figures are averaged in $y$ since there is no significant variation in that direction. Fig.~\ref{Te} shows the time evolution of electron temperature in the region where the hot and cold electrons come into contact. The initial profile (black) represents a high temperature region decreasing to a low temperature region through a narrow transition that has a length scale of several $d_e$'s. We do not expect the results to be sensitive to this transition scale since we find that the DL develops from a streaming instability after the temperature discontinuity has broadened. Over a short period of time ($\sim$100 $\omega_{pe}^{-1}$, blue), the transition broadens as the hot electrons free-stream into the cold electron region. If free-streaming were to continue, we would expect the profile to continue to broaden due to the mixing of the two populations. However, a small jump develops (see $\omega_{pe}$t=300 ) within the smooth transition, meaning that the escape of hot electrons by free-streaming is inhibited. Over time, the jump grows in size and significantly suppresses mixing due to free-streaming. As a result, the previously smooth transition is converted into a distinctive jump across two separate temperature regions. This indicates that the transport of hot electrons into the surrounding plasma is being suppressed. Other features of Fig.~\ref{Te} include Langmuir waves propagating in the cold electron side (see e.g., the wiggles on the right side of the curve at $\omega_{pe}$t=300) that are excited by the bump-on-tail instability as hot electrons free-stream into that region. Beam modes are excited on the hot electron side (see, e.g., longer scale waves on top of the curve at $\omega_{pe}$t=1400) by the electron-electron streaming instability as cold electron beams that are accelerated by a DL enter the hot side and interact with the hot electrons.

\begin{figure*}
\centering
\subfigure[]
{\label{Ex_timehistory}
\includegraphics[scale=0.6]{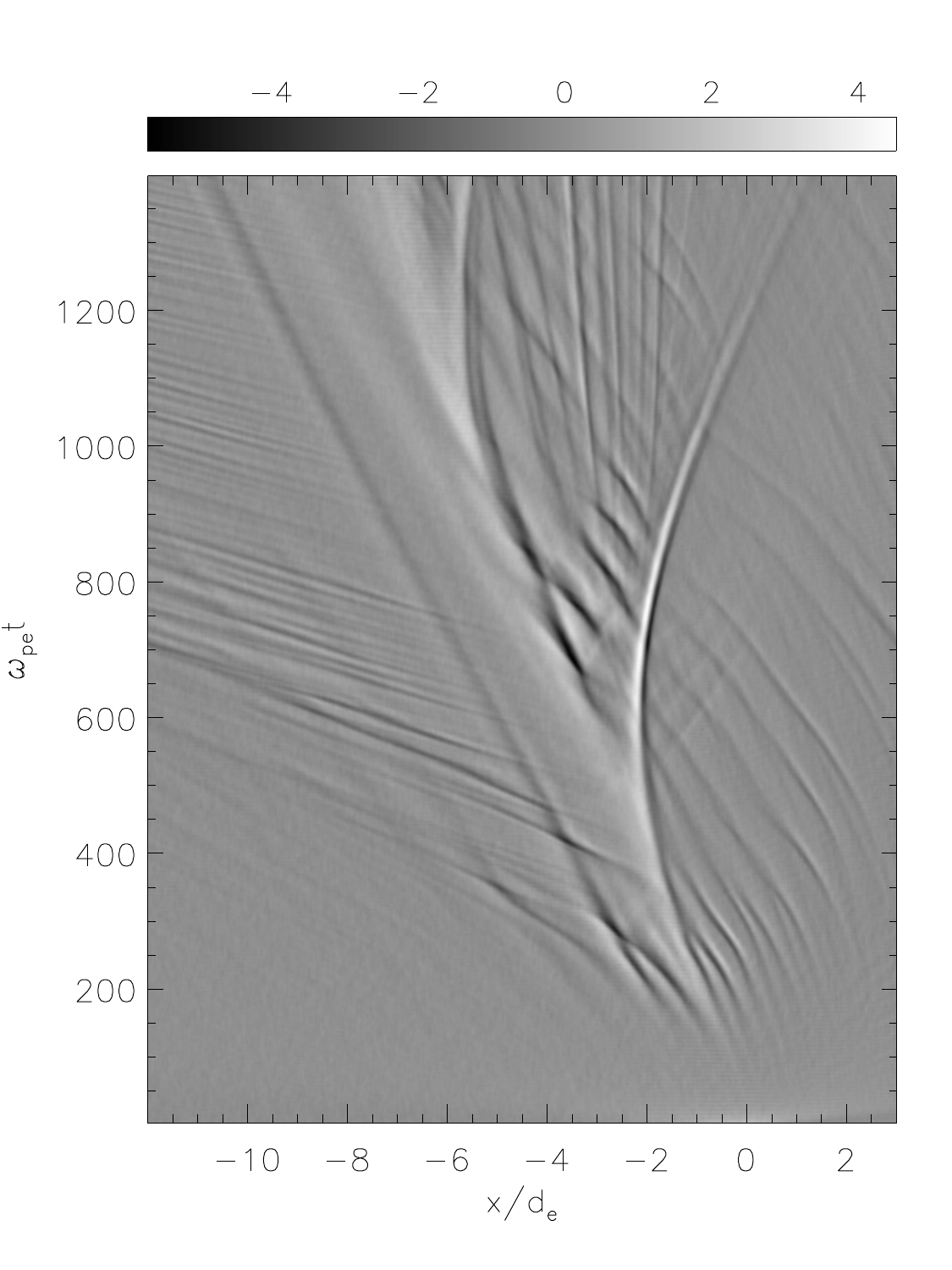}
}
\subfigure[]{\label{phiRE}
\includegraphics[scale=0.6]{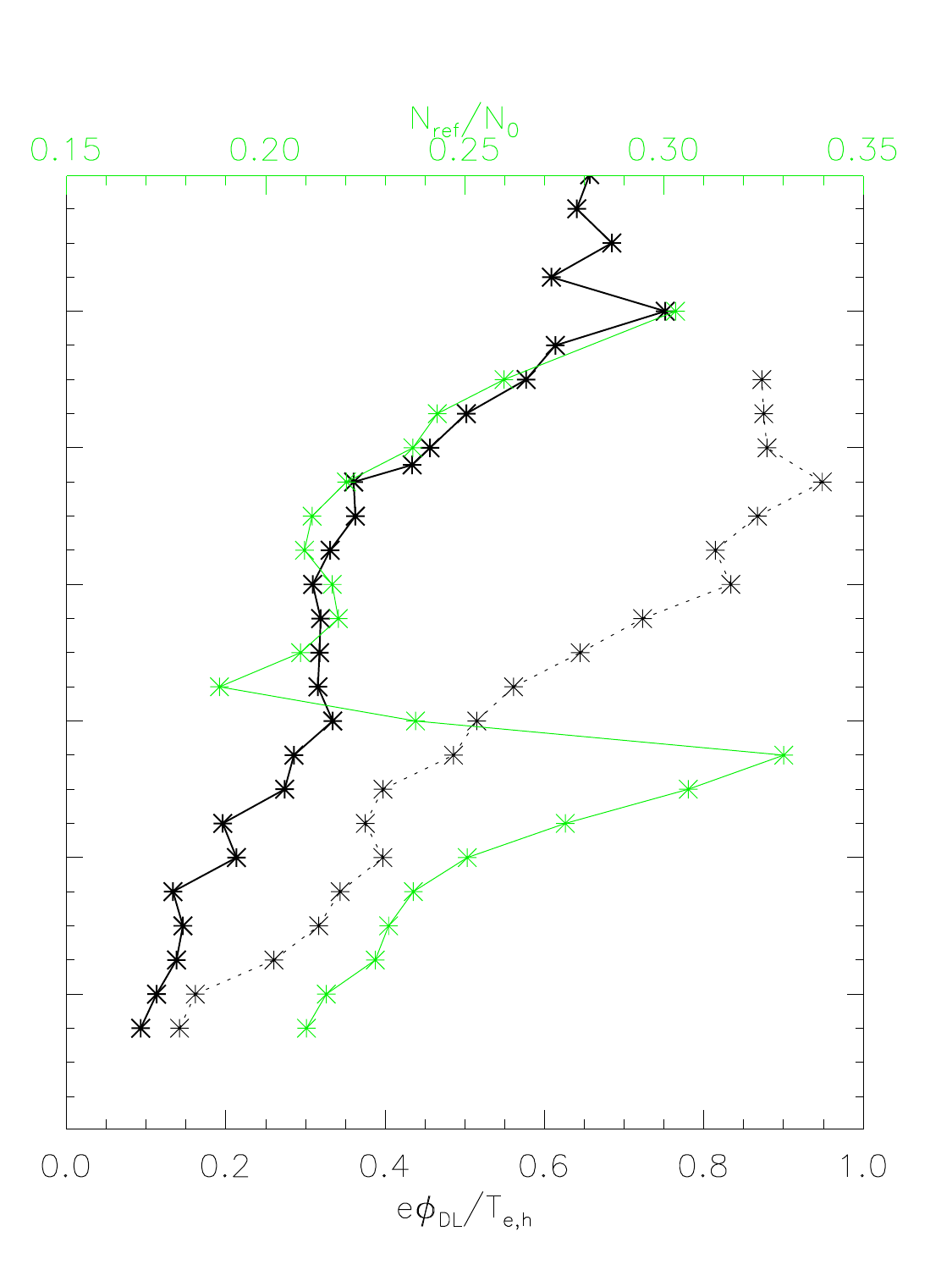}
}

\centering
\caption{\label{Ex_phi} Evolution of (a) $E_x$, the parallel electric field and (b) $e\phi_{DL}/T_{e,h}$ (solid black), the electric potential jump across the DL normalized to the instantaneous hot electron temperature at the center of the hot region (dotted line is from a second simulation with $T_{e0,h}$=1 that will be discussed later), overlaid with $N_{ref}/N_0$ (solid green), the fraction of return current electrons reflected at the foot of the DL divided by the total initial electron number.   }
\end{figure*}

\begin{figure}
\includegraphics[scale=0.52]{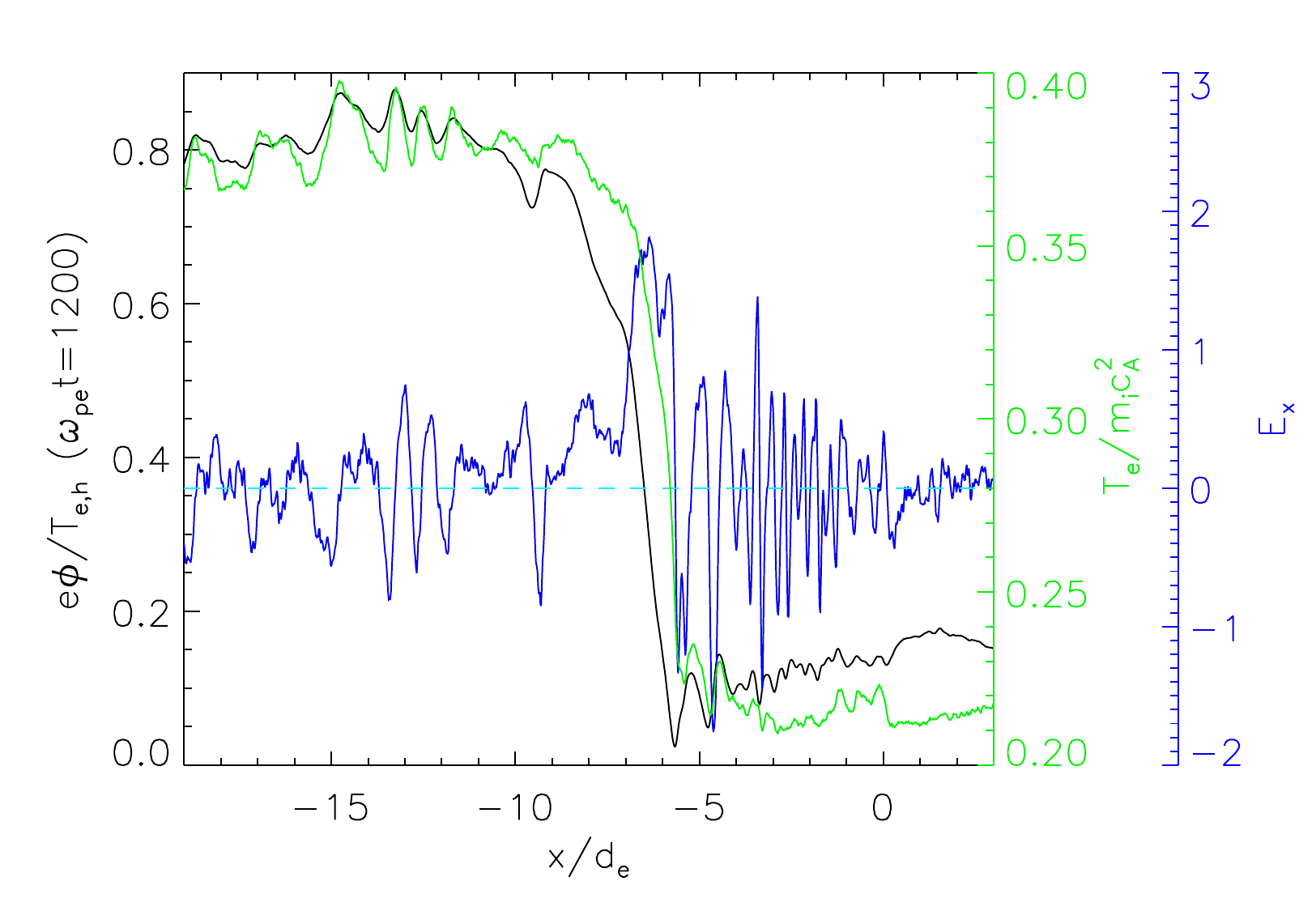}
\caption{  \label{oplot1200} $e\phi/T_{e,h}$ (black) and $T_e$ (green), overlaid with $E_x$ (blue) and the zero $E_x$ position (dotted cyan), at $\omega_{pe}$t=1200. The DL is characterized by a monotonic drop in the potential. The potential barrier suppresses the transport of hot electrons and results in a drop in $T_e$ across the DL. }
\end{figure}

\subsection{DL evolution and strength}

The heat flux suppression comes from a DL that arises within the transition region. Figure~\ref{Ex_timehistory} shows the time history of the electric field parallel to the local magnetic field, $E_x$, and \ref{phiRE} the time evolution of the electric potential jump across the DL normalized to the instantaneous hot electron temperature measured at the center of the hot region, $e\phi_{DL}/T_{e,h}$. $T_{e,h}$ in $e\phi_{DL}/T_{e,h}$ thus represents the core temperature of the hot electron population. $T_{e,h}$ drops over time as can be seen on the left side of Fig. \ref{Te} because hot electrons continuously leak out. We normalize the potential jump of the DL to $T_{e,h}$ to more accurately reflect the suppression strength of the DL on the hot electrons that remain in the source region at any particular time. Also shown in Fig. \ref{phiRE} is the fraction of return current electrons that are reflected at the foot of the DL, normalized to the initial total electron number, $N_{ref}/N_0$. The reflection mechanism and the reason for its importance are discussed later. $E_x$ is averaged over an electron plasma period to eliminate initial fluctuations in the contact region that decay over time. The DL emerges at $\omega_{pe}$t$\sim$100 around x=0. At $\omega_{pe}$t$\sim$550, it emits a large amplitude shock that moves in the positive x direction. The DL is the more structured dark region in Fig.~\ref{Ex_timehistory} that slowly drifts in the negative x direction. The DL is constantly evolving over the entire course of the simulation. At $\omega_{pe}$t=1200, it reaches its maximum amplitude and extends over a scale of 4 $d_e$, centered at x$\sim$ -6$d_e$. Fig.~\ref{oplot1200} is a cut of Fig.~\ref{Ex_timehistory} at this time in a blowup around its location. Shown are $E_x$ (blue), the electric potential normalized to the core hot electron temperature $e\phi/T_{e,h}$ (black) and the electron temperature $T_e$ (green). $E_x$ has a large positive peak at x$\sim$ -6$d_e$ that causes a large potential drop that reflects hot electrons and produces a sharp drop in the electron temperature. We define the strength of the DL by its potential jump $e\phi_{DL}$ which is calculated as follows. To the left of the DL in Fig.~\ref{oplot1200}, the potential increases, oscillates and eventually reaches a fairly constant value (near x$\sim$ -13$d_e$), the high end value. To the right of the DL, the potential drops and oscillates about a roughly constant value (near x$\sim$ -5$d_e$), the low end value. $e\phi_{DL}$ is the difference between the high and low end values and is shown as a function of time (solid black) in Fig. \ref{phiRE}. $e\phi_{DL}/T_{e,h}$ is a measure of the suppression capability of the DL on the hot electron transport. If the DL is large enough such that $e\phi_{DL}/T_{e,h}\sim$ 1, the thermal bulk of the hot electrons will be reflected by the potential barrier of the DL. This will imply a significant reduction of the electron heat flux as well.

\begin{figure}
\includegraphics[scale=0.55]{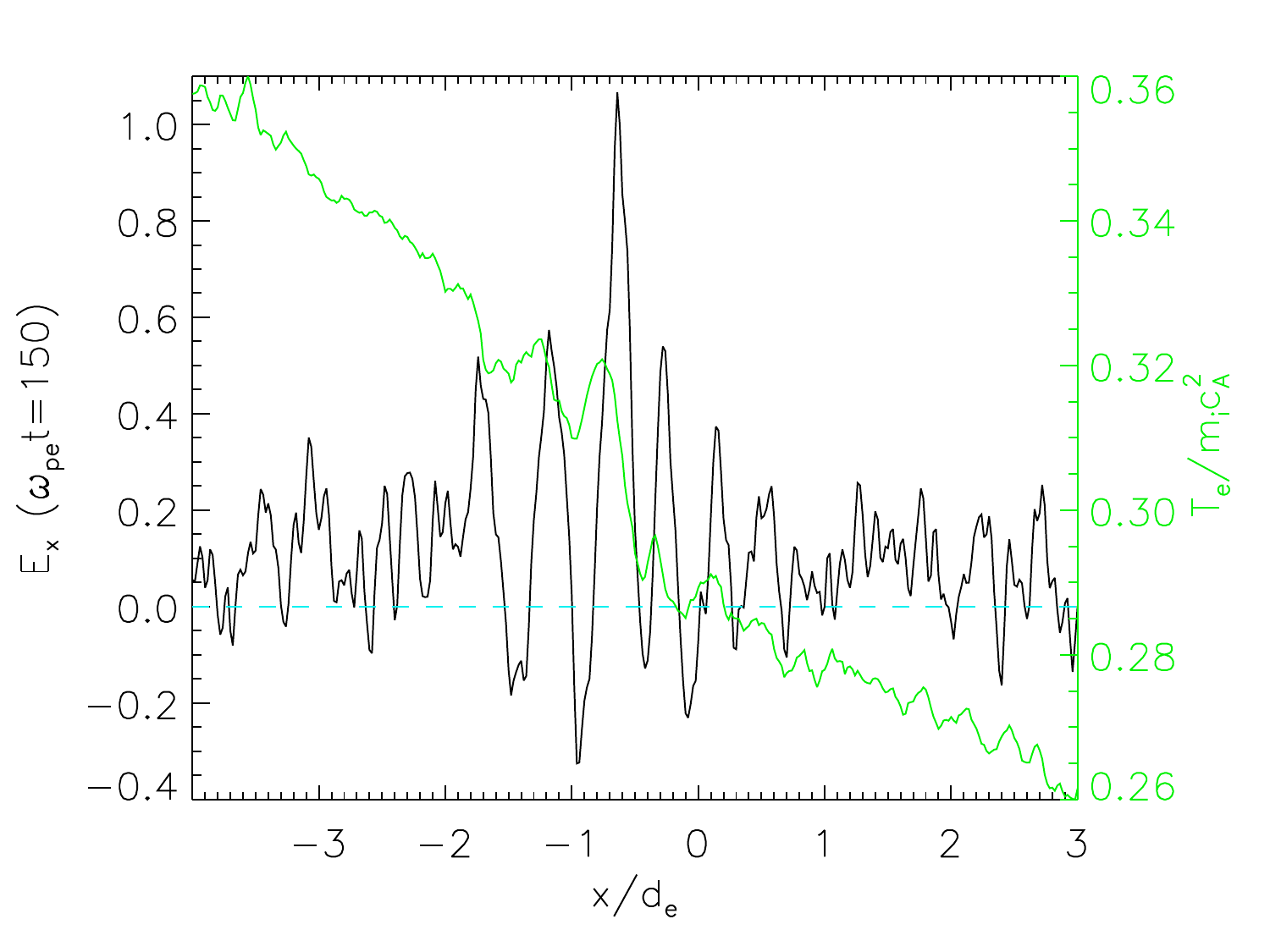}
\caption{ \label{oplot150}  The electric field $E_x$ at $\omega_{pe}$t=150 , overlaid with $T_e$ (green). The large amplitude wave that forms around x$\lesssim$0 arises from an ion/return-current-electron streaming instability.  }
\end{figure}

Note also in Fig.~\ref{oplot1200} that the DL is not merely a unipolar electric field. It has a small negative leg between x= -6 and -5$d_e$, on the low $T_e$ side. While the bigger positive part suppresses streaming hot electrons coming from the left, the negative leg can reflect cold electrons coming from the right (low $T_e$ side). The reflection of these electrons is the drive mechanism of the DL, which we now discuss. Fig.~\ref{oplot150} is a cut of Fig.~\ref{Ex_timehistory} at $\omega_{pe}$t=150, shortly after the DL emergence. Multiple peaks in $E_x$ develop in the $T_e$ transition region. In the linear regime ($\omega_{pe}$t$<$150), these waves have small amplitude, comparable to the background fluctuations. They gradually grow into large-amplitude nonlinear structures at later times. The peak which has the largest amplitude dominates the others and becomes the DL. It then starts to hinder hot electron free-streaming and produces an obvious drop in $T_e$ as can be seen at $\omega_{pe}$t=150. The two smaller peaks on either side of x=0 later turn into fairly symmetric waves, which produce no significant potential jump. The wave turbulence from which the DL emerges is driven by the return current interacting with the background ions. The DL is also driven by this streaming instability, whose origin is now described.

\subsection{DL drive mechanism: streaming instability}

\begin{figure*}
\centering
\subfigure[]
{\label{phase150}
\includegraphics[scale=0.6]{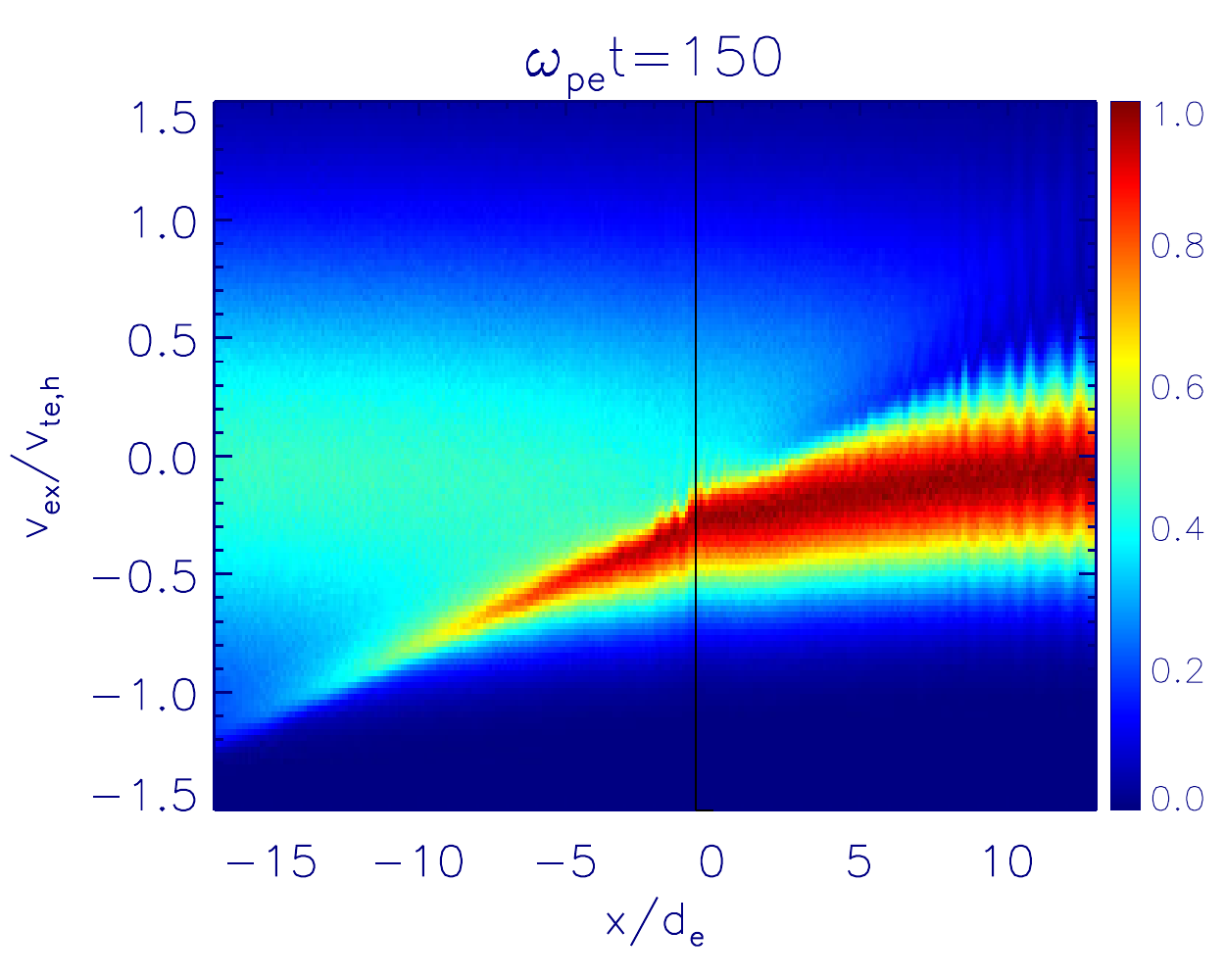}
} 
\subfigure[]
{\label{cut150}
\includegraphics[scale=0.6]{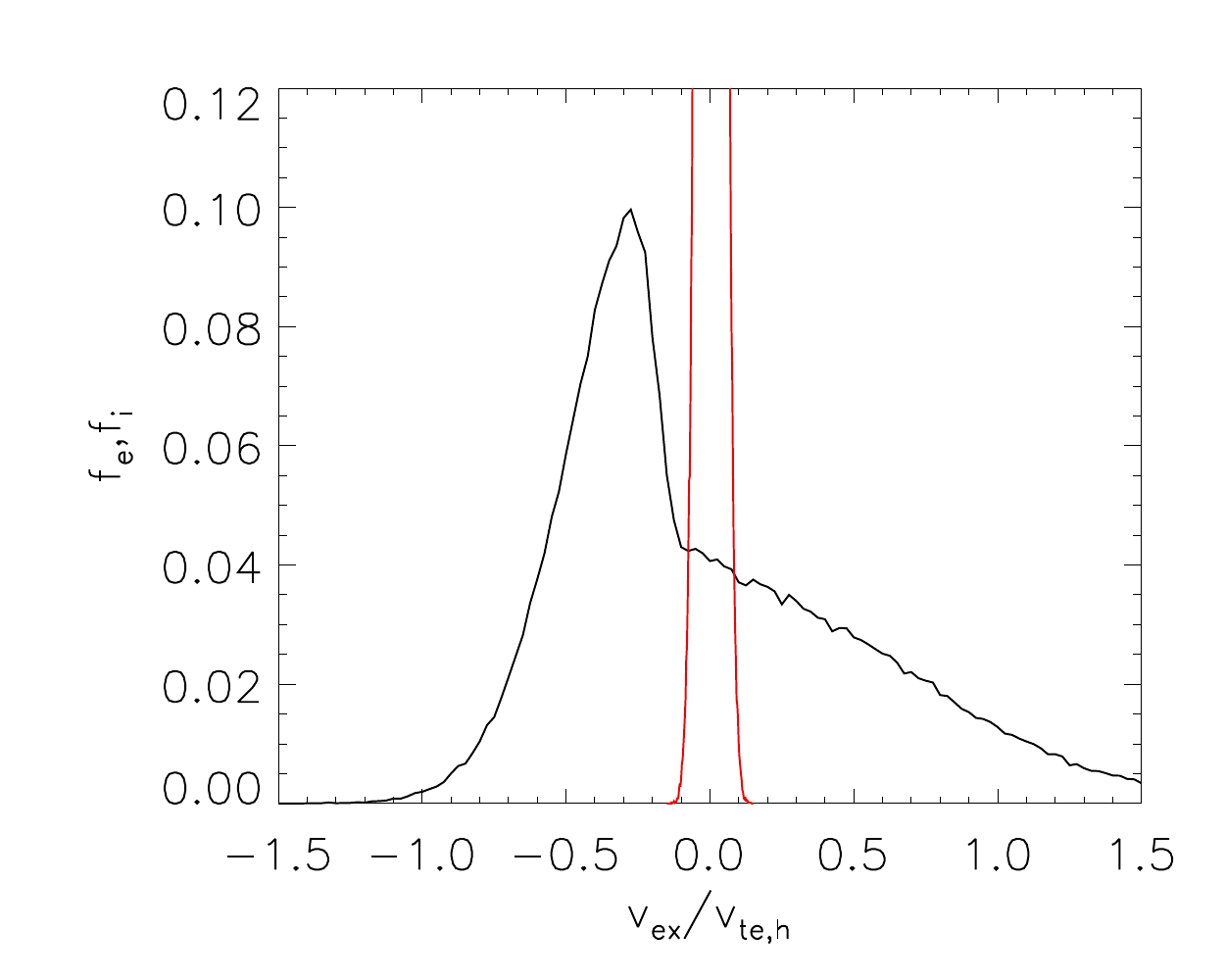}
}
\centering
\caption{\label{phasencut150} (a) Electron phase space at $\omega_{pe}$t=150 and (b) a cut of (a) at the position of the DL at x$\sim$ -1$d_e$, indicated by a vertical black line. The ion distribution $f_i$ (red) is overlaid on the electron distribution $f_e$ (black). Since the ions and cold electrons have the same temperatures, the ion thermal speed is $\sim\sqrt{m_i/m_e}$=10 times lower than the return current thermal speed. The integrated areas under the curves (i.e., their densities) are the same, so $f_i$ peaks at $\sim$10 times the height of $f_e$. The vertical scale is normalized to the maximum of $f_i$.    }
\end{figure*}

Figure~\ref{phase150} shows the electron phase space at $\omega_{pe}$t=150. The hot and cold electron distributions are at the left and right sides, respectively. Note the narrower thermal spread of the cold electrons at the right, as well as the wider spread of hot electrons at the left. The initial temperature transition is near x=0. A beam of cold electrons is entering the hot region. At $\omega_{pe}$t=0, both distributions are non-drifting and there is no beam. As the hot electrons free-stream into the cold electron region, ions cannot follow and are left behind as a charge imbalance. The resulting electric field drags a cold electron beam towards the hot electron region as a return current. The relative drift between this return current and the ions excites the instability. Fig.~\ref{cut150} is a cut of the phase space at the position of the DL (x$\sim$ -1$d_e$). The return current, with a negative beam speed, sits on top of the streaming hot distribution, which is moving to the right and has a wider thermal spread. The drift of the return current electrons with respect to the ions (red) excites the ion/return-current-electron streaming instability. The linear growth rate $\gamma$ of the ion-electron streaming instability (also known as the Buneman instability \citep{Buneman58}) for the most unstable wave calculated in fluid theory is given by
\begin{equation}
\gamma = \frac{\sqrt{3}}{2}\left(\frac{m_{e}}{2m_{i}}\right)^{1/3} \omega_{pe}.\label{eqn1}
\end{equation}
Using the parameters of our simulation, $\gamma\sim$ 0.15 $\omega_{pe}$. The growth time $\tau$ is thus $\tau\sim\gamma^{-1}\sim$7 $\omega_{pe}^{-1}$. The instability is sufficiently strong to produce the DL at $\omega_{pe}$t$\sim$100.

\begin{figure*}
\centering
\subfigure[]
{\label{phase550}
\includegraphics[scale=0.6]{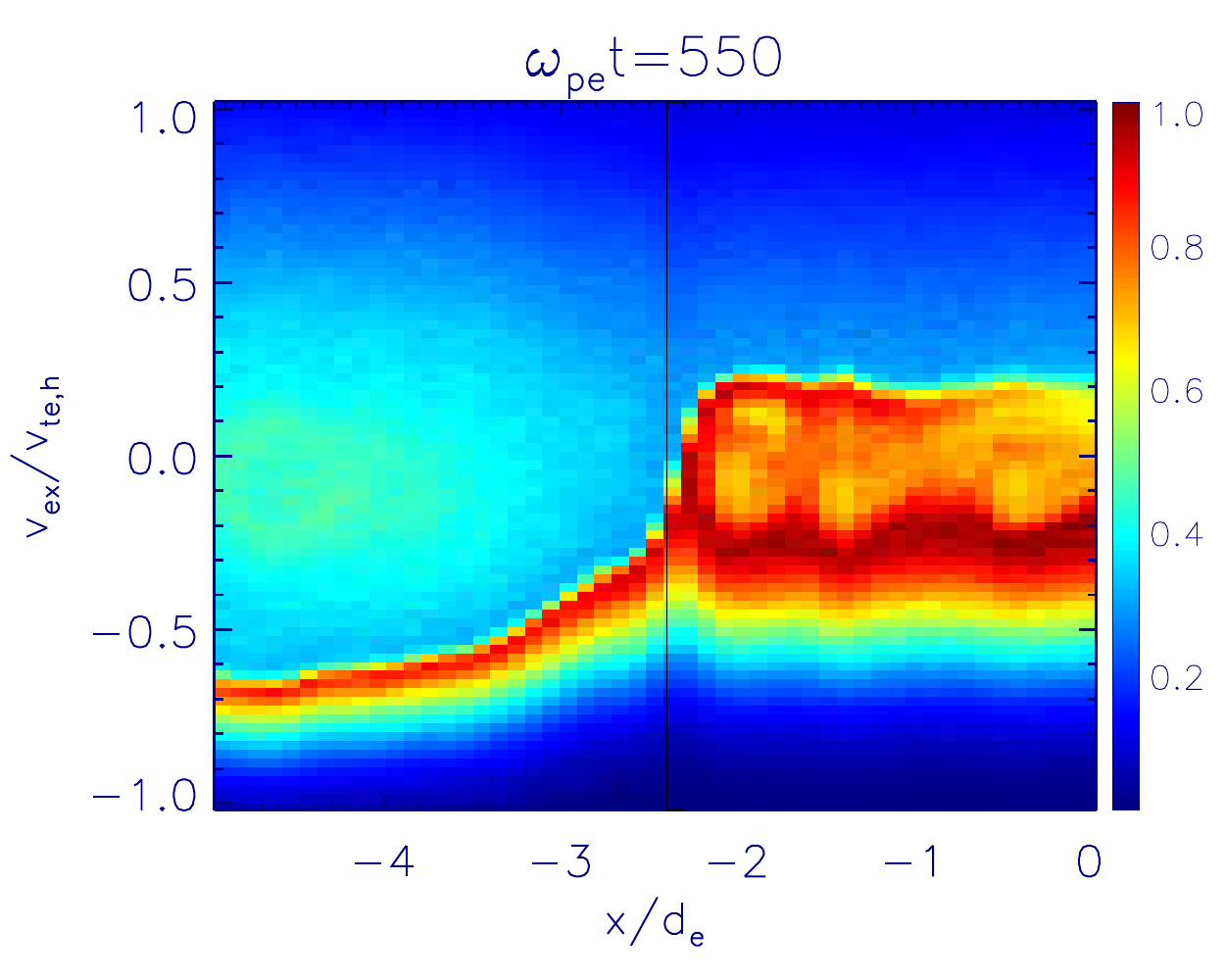}
}
\subfigure[]
{\label{phase700}
\includegraphics[scale=0.6]{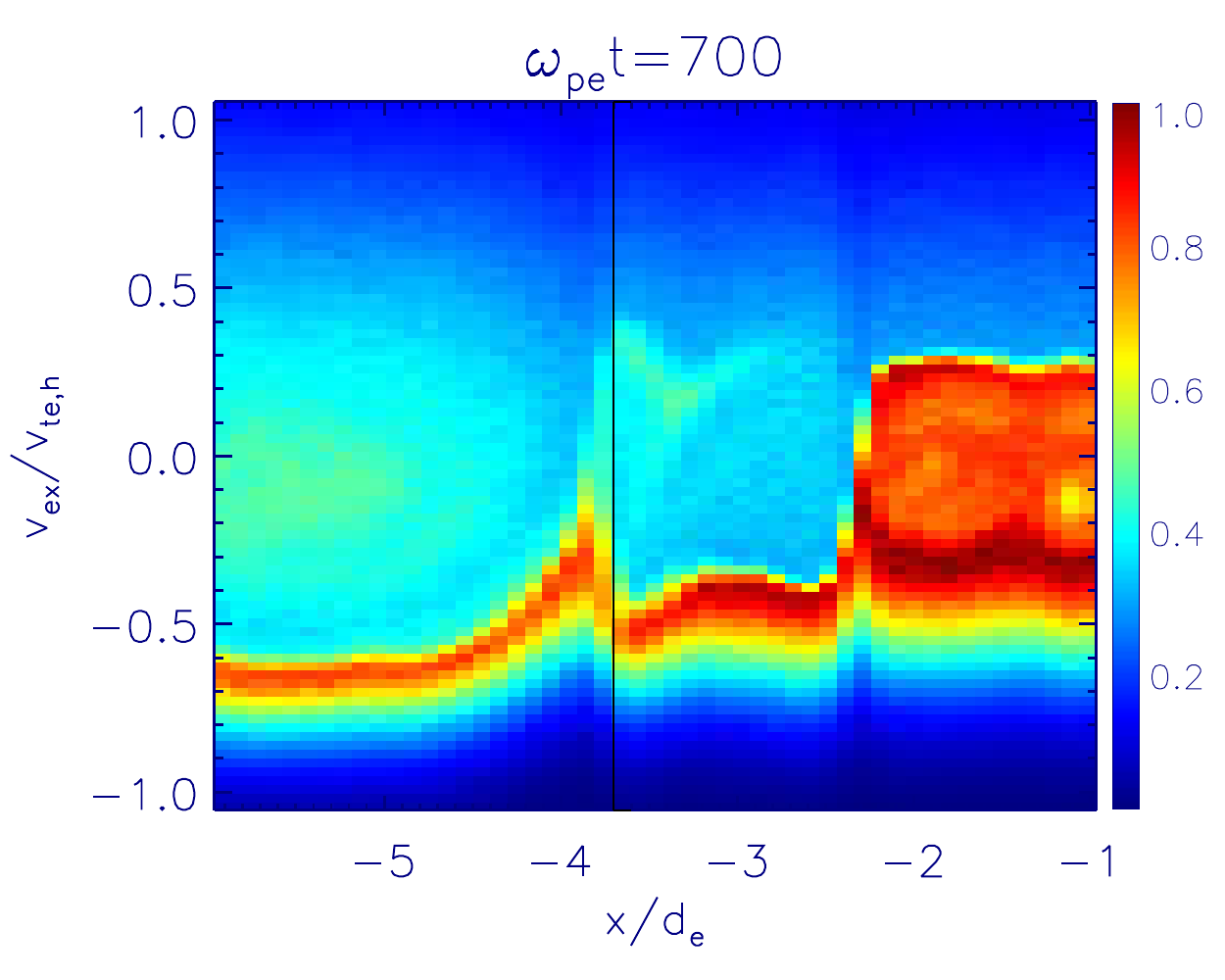}
}
\centering
\caption{\label{phase550700} Electron phase spaces at: (a) $\omega_{pe}$t=550; and (b) $\omega_{pe}$t=700. The vertical black lines mark the low potential side of the DL and to its right is the negative leg $E_x$ of the DL that can reflect the return current electron beam. }
\end{figure*}

The time evolution of the electron phase space reveals that the growth of the DL depends on the number of return current (RC) electrons reflected at the foot of the DL. One can see from Fig.~\ref{oplot1200} that the electric field $E_x$ of the DL has a negative leg on the low $T_e$ side. This shows up as a potential dip on the low potential side. The RC electrons are reflected by the negative leg, which effectively reduces the RC speed and thus converts the kinetic energy of the RC to the electrostatic energy of the DL. By exchanging momentum with the reflected electrons, the DL gets amplified \citep{Lotko83}. Electron reflection also indicates when the instability enters into a nonlinear regime ($\omega_{pe}$t$>$150). The DL growth depends on the number of RC electrons reflected at the foot of the DL. Qualitatively, the dependence of DL growth on the number of reflected electrons can be seen by contrasting the phase space at a time when the DL is growing, e.g., at $\omega_{pe}$t=550, with a time when it is not growing, e.g., at $\omega_{pe}$t=700 (as $e\phi_{DL}/T_{e,h}$ plateaus in Fig.~\ref{phiRE}). At $\omega_{pe}$t=700 (Fig.~\ref{phase700}) to the right of the DL at x$\gtrsim$ -4$d_e$ (vertical black line), there are many fewer reflected electrons than to the right of the DL at $\omega_{pe}$t=550 (Fig.~\ref{phase550}), x$\gtrsim$ -2$d_e $. We estimated the number of reflected RC electrons, $N_{ref}$, by integrating the electron distribution to the right of the DL over a distance of one $d_e$, which is the typical width of the DL, and for $v_{ex}\in$(0, 0.4 $v_{te,h}$), which is approximately the velocity range covered by the reflected RC electrons as they turn around and acquire positive velocities. $N_{ref}$ is normalized to $N_0$, the initial electron number contained in the same interval. Note that the integration for $N_{ref}$ also includes a contribution from hot free-streaming population in that velocity range. However, this contribution stays nearly constant and the variation in $N_{ref}$ mainly comes from the change in the number of reflected cold RC electrons. Therefore, $N_{ref}$ suffices as an estimate of the reflected RC electrons (plus some constant offset). $N_{ref}/N_0$ (green) is plotted on top of the DL size, $e\phi_{DL}/T_{e,h}$, in Fig.~\ref{phiRE} to show their correlation. As $N_{ref}/N_0$ increases during $\omega_{pe}$t$\leq$550 and $\geq$950, $e\phi_{DL}/T_{e,h}$ has a general increasing tendency as well. As $N_{ref}$ drops abruptly at $\omega_{pe}$t=550 and stays at a low level until $\omega_{pe}$t=950, $e\phi_{DL}/T_{e,h}$ plateaus. Hence, there is good correlation between the DL growth and the number of the reflected RC electrons.

An examination of the evolution of the electric field in Fig.~\ref{Ex_timehistory} reveals that at $\omega_{pe}$t$\sim$550, the DL sheds an ion-acoustic shock wave, which shows up as an ion hole in ion phase space. Once it leaves the DL, the shock accelerates to the right while the DL continues to slowly move to the left. Therefore, their separation increases. The shock wave is bipolar and reflects most of the RC electrons with its large-amplitude negative leg. One can see in Fig.~\ref{phase700}, at $\omega_{pe}$t=700, most of the RC electrons are reflected behind the shock at x$\lesssim$ -2$d_e$. $N_{ref}/N_0$ shows an abrupt drop at $\omega_{pe}$t=550 because most of the RC electrons are stuck behind the shock and do not reach the DL. The shock dissipates through ion heating. The ion temperature increases by as much as a factor of two along the path traversed by the shock. The shock gradually diminishes in amplitude. Similar dynamics (acceleration and decay) of an ion hole mode at the low potential side of a weak DL created in ion-acoustic turbulence is also observed in current-driven systems \citep{Barnes85}.

\begin{figure}
\includegraphics[scale=0.6]{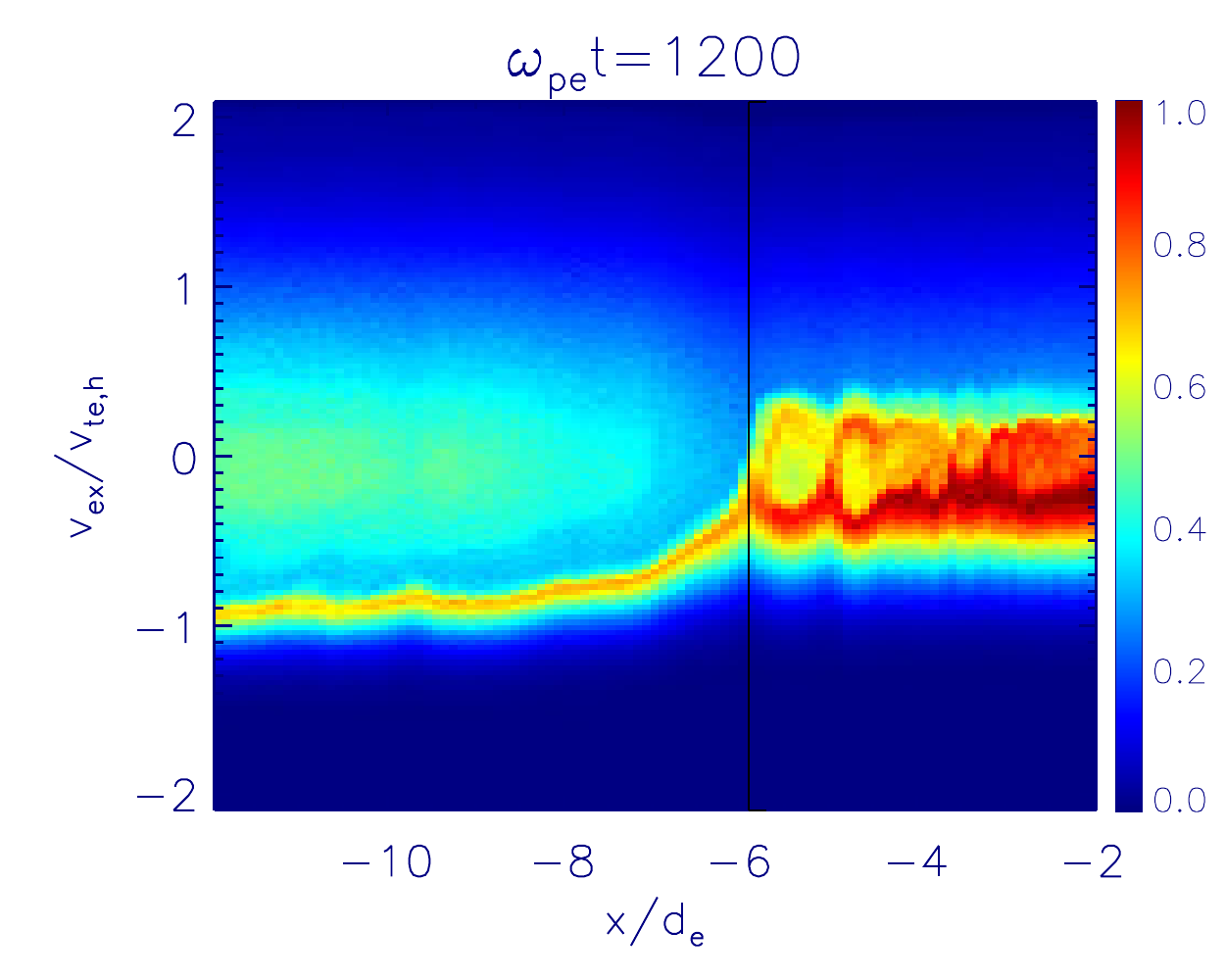}
\caption{ \label{phase1200}  Electron phase space at $\omega_{pe}$t=1200 with $e\phi_{DL}/T_{e,h}\sim$0.8. To the left of the DL location, x$\sim$ -6$d_e$, hot electron populations with $v/v_{te,h} <$ 0.9 (cyan and green in color) are reflected by the potential barrier. They are trapped in the hot region. For higher velocity electrons (at the tail; blue in color), there is a drop in velocity as they are decelerated. Both contribute to suppression of the heat flux.  }
\end{figure}

During our simulation, $e\phi_{DL}/T_{e,h}$ reaches 0.8 at $\omega_{pe}$t=1200 (Fig.~\ref{oplot1200}). This means that hot electrons with velocities $v/v_{te,h} \leq \sqrt{\mathrm{0.8}} \sim$ 0.9 will be reflected. The electron phase space at $\omega_{pe}$t=1200 in Fig.~\ref{phase1200} reveals that the population of hot electrons with $v/v_{te,h} \leq$ 0.9 wraps around from positive to negative velocities when encountering the DL. So they are reflected and hence confined in the source region. The suppression observed here is significant. We calculate the heat flux carried by kinetic hot electrons that is positively directed as $Q(x,t)$=$ \frac{1}{2}m_e\int\limits_0^{\infty}v^3f_e(x,v,t)dv$ \citep{Krall73}. The mean flow speed is $<$2\% of the hot electron thermal speed and hence negligible in the calculation of $Q(x,t)$. Similar to $T_e$, $Q$ has a sharp jump across the DL. Towards the end of the simulation, $Q$ is suppressed by $\sim$40\% to the right of the DL compared with the higher value to the left of the DL.

\subsection{Dependence of DL strength on hot electron temperature}

To determine the dependence of the DL potential drop on the hot electron temperature, we performed another run in which we raised the initial hot electron temperature $T_{e,h,\parallel}$ from 0.5 to 1, and $B_0$ from 1 to $\sqrt{2}$ such that $\beta_{e,h,\parallel}$ remains unity, while keeping all other parameters unchanged. Slightly increasing $B_0$ should not have an effect on the result because the phenomena we observed here are dominantly electrostatic. 

The size of the DL depends on the hot electron temperature. The DL potential drop $e\phi_{DL}/T_{e,h}$ from another simulation with a higher initial hot electron temperature $T_{e0,h}$=1 is plotted as a dotted line in Fig.~\ref{phiRE}. This potential drop is normalized to the instantaneous hot electron temperature at the center of this new hot region. The DL is on average stronger than that in the simulation with initial $T_{e0,h}$=0.5 (solid line). The basic dynamics of the DLs are the same in both cases. A phase of growth is followed by a period of pause, and the whole cycle repeats. The DL evolves and grows faster in the higher $T_{e0,h}$ run, probably because the return current electron response, which is the driver of the DL, is faster due to the increased thermal speed of the hot electrons (relative to those of the ambient plasma) in the $T_{e0,h}$=1 run than in the $T_{e0,h}$=0.5 run.  There is also a modest increase in the suppression of heat flux $Q(x,t)$ from $\sim$40\% in the $T_{e0,h}$=0.5 run to $\sim$50\% in the higher $T_{e0,h}$ run. Therefore, one might expect a higher ratio of hot to cold electron temperatures would lead to a stronger DL and associated suppression of transport.

\section{\label{dis_app}DISCUSSION AND APPLICATION}

We have studied the suppression of electron thermal transport due to double layer (DL) formation in 2D PIC simulations. However, our system is effectively 1D since the transverse direction is only $\approx$ 20$\lambda_{De}$ wide. The presence of the second dimension enables us to average the data for noise reduction. 2D effects on DLs are not investigated in this study, but do not seem to be significant. Extended (planar) weak DLs over 80 $\lambda_{De}$ in transverse extent were observed in earlier 2D PIC simulations where subthermal electron current was injected at the boundaries of the parallel direction \citep{Barnes85}. Such electron injection may correspond to the return current electrons being drawn in at the contact region of the two electron populations in the present simulations. The planar 2D DLs tend to develop substantial substructure across the magnetic field late in time, but the basic dynamics of the DLs are the same as in 1D. We expect the formation of DLs, and therefore transport suppression of hot electrons, to persist in 2D. However, 2D DLs are unlikely to be planar over $\sim$ 10$^8\lambda_{De}$, the scale of coronal looptops. On the other hand, a variation of the DL over a large transverse distance is not likely to strongly impact the largely 1D dynamics studied here. The injection of a narrow band of hot electrons, however, might behave very differently. This case remains to be studied.

We also note that the computational size of the PIC simulations, which is 655$d_e\sim$ 66m in the corona, is small compared to the actual size of flaring loops. They, however, reveal important physics down to the kinetic electron scale, the Debye length, where DLs form. We expect the DL to continue strengthening in a larger domain that allows the system to evolve for a longer period. 


Applying our results to the above-the-looptop hard X-ray sources \citep{Krucker10, Krucker07} that show an exponential decay on a timescale of more than two orders of magnitude longer than the electron transit time through the source, the formation of a DL provides a plausible mechanism to confine energetic electrons in the source region. Electrons that have kinetic energies less than the potential drop of the DL are reflected back to the source region. Over the entire course of the simulations, the DL tends to amplify through multiple phases of growth. There are no signs of decay. Since it is driven by the return current electrons, which exist for as long as some hot electrons are leaking out, the DL can persist until the eventual depletion of all hot electrons in the source region, which should take much longer than the electron transit time through the source due to the strong transport suppression caused by the DL. The DL lifetime is therefore expected to be much longer than the transit time. This can naturally explain the prolonged lifetimes ($>$100 s) of the energetic electrons from the above-the-looptop hard X-ray observations compared with the electron transit time ($\sim$1 s) through the source.

\section{\label{con}CONCLUSION}

Results from PIC simulations of a pre-accelerated hot electron source in contact with ambient plasma, which provide a basic model of coronal looptop sources, are presented. The hot electrons free-stream along the magnetic field initially, but transport is suppressed by the formation of a double layer that is driven by an ion/return-current-electron streaming instability due to the drift of return current electrons with respect to the ions. The DL subsequently reflects and decelerates hot electrons as they begin to propagate into the surrounding plasma. A significant reduction in electron heat flux through the ambient electron region is observed. Such a suppression mechanism differs from the conventional picture of electron scattering by turbulence. Our results are consistent with observations of both free-streaming propagation \citep{Aschwanden96} and the suggested confinement at looptop \citep{Masuda94, Krucker10, Krucker07} of hard X-ray producing electrons, corresponding to the escaping and reflected hot electrons, respectively.


\bibliographystyle{astroads}




\end{document}